\documentclass[aps,prd,twocolumn,superscriptaddress]{revtex4}
\usepackage{graphicx}
\usepackage{bm}
\usepackage{amsmath,mathrsfs}
\usepackage{amssymb}
\usepackage{slashed}
\usepackage{latexsym}
\usepackage{epsfig,subfigure}
\usepackage{amsbsy}
\usepackage{array}
\usepackage{amssymb}
\usepackage{changes}
\usepackage{color}
\usepackage{setspace}
\usepackage{lipsum}
\usepackage{float}
\usepackage{color}
\usepackage{changes}
\usepackage[T1]{fontenc}
\usepackage{mathptmx}
\usepackage{diagbox}
\DeclareMathAlphabet{\mathcal}{OMS}{cmsy}{m}{n}
\DeclareSymbolFont{largesymbols}{OMX}{cmex}{m}{n}
\usepackage{CJK}

\begin{document}

\title{     The chiral phase transition and equation of state in the chiral imbalance }

\author{Qing-Wu Wang }
 \email{qw.wang@scu.edu.cn }
 \affiliation{College of Physics, Sichuan University, Chengdu 610064, China}

  \author{Chao Shi}
  \email{ shichao0820@gmail.com}
  \affiliation{Department of nuclear science and technology,
Nanjing University of Aeronautics and Astronautics, Nanjing 210016, China;}

\author{Hong-Shi Zong}
\email{zonghs@nju.edu.cn}
\affiliation{Department of Physics, Nanjing University, Nanjing 210093, China}
\affiliation{Department of Physics, Anhui Normal University, Wuhu, Anhui 241000, China}
\affiliation{Nanjing Institute of Proton Source  Technology  , Nanjing 210046}

\begin{abstract}
The chiral phase transition and equation of state are studied within a new self-consistent mean-field approximation of the two-flavor Nambu$-$Jona-Lasinio model. In this newly developed  model, modifications to the   chemical  potential $\mu$ and chiral chemical potential $\mu_5$ is naturally included by adding vector and axial-vector channels from  Fierz-transformed Lagrangian to the standard Lagrangian.
In proper-time scheme, the chiral phase transition is a crossover in the $T-\mu$ plane. But when    $\mu_5$  is increased, our study shows that there may exist first order phase transition. Furthermore, the  chiral imbalance will soften the equation of state of quark matter. The mass-radius relations and tidal deformability of  quark stars are calculated. As $\mu_5$ increases, the maximum mass and radius  decrease. The vector channel and axial-vector channel have opposite influence on the equation of state.  However, when EOS is constrained by astronomical observations, the shape of the mass-radius curve can be used to determine whether there is chiral imbalance in the dense object, and thus indirectly proving the CP violation in the dense matter.  Our study shows a different influence  of the chiral imbalance  on  the chiral phase  transition in contrary to tree-momentum-cutoff scheme.


\end{abstract}

\maketitle
\section{Introduction}

The phase diagram of strongly interacting matter is an important topic in hadron physics. Under extreme conditions, the hadron state will undergo phase transition from hadron to quark  and the restoration of spontaneous chiral symmetry \cite{Wilczek,Buballa,Fukushima1,Luo}. At finite temperature and finite density, the chiral restoration may be a first-order phase transition, but the results of different models and even different regularization are inconsistent.  At zero baryon density, the lattice Monte Carlo simulations give   reliable result that the chiral transition is a crossover, but when the temperature is zero and the density is high, the lattice calculation faces the sign problem \cite{Fodor,Gavai}.

  At high density,   topological gauge fields  with nonzero winding number (instantons and  sphalerons) may appear  \cite{Hooft,Witten,Chu,Gross,Schafer,Arnold}.
  The interaction of  quarks  with these topological gauge fields  will change the helicities of the quarks, which results in the
chiral imbalance between left- and right-hand quarks via the axial anomaly \cite{Adler,Christ,Smilga}.   The interaction of  quarks  with these topological gauge fields would also lead to a local $P$ and $CP$ violation.
 The strong magnetic
fields are suggested to be produced at the very first moments of a noncentral heavy ion collision \cite{Kharzeev1,Kharzeev2}. If the chiral imbalance is obvious here, it will result in the observable effects in experiment since the right- and left-hand quarks move in different directions along the magnetic field.
   This phenomenon is called as chiral magnetic effect (CME) \cite{Fukushima} and can be served as  indirect evidence for   $P$ and $CP$ violation \cite{Kharzeev3,Kharzeev4}.

The  chiral imbalance  means an asymmetry in
the number of right- and left-handed quarks which.   In order to study the effect of this asymmetry,  a chiral chemical potential  $\mu_5$ that conjugated  to  the  chiral  charge  density $n_5$, can be introduced \cite{ Fukushima,Ruggieri}. As showed in Refs. \cite{Ruggieri,Yang,Luya}, this chiral chemical potential affect the position of critical end point (CEP).   Furthermore, on the theoretical side, the exist and location of CEP depend on models and regularisation.
  To confirm and find the  CEP  is a hot issue \cite{Asakawa,Qin,Fischer}. Even if CEP exists, its location is still uncertain.
   People expect to give relevant information from experiments \cite{Hatta,Lacey,Adamczyk,Abelev} and astronomy \cite{Most,Orsaria,Hanauske,Bauswein}.




Before the study of chiral phase transition, an  appropriate  regularisation scheme must be chosen. The CEP exists in the three-momentum cutoff  scheme but disappears in the proper-time regularisation scheme.
  So it is wonder whether the CEP exists  in a  chiral imbalance system. We will exam if the existence of chiral imbalance  could lead to the chiral phase transition from crossover to a first-order transition. If the first-order phase transition is found in QGP but without properly regulation of the chirally imbalance, it is still uncertain whether the first-order transition is caused by high density or chiral imbalance.

Since the chiral imbalance  changes the possible location of CEP, it will naturally affect the equation of state (EOS). However,  because the  quantum anomaly in defining chiral density, $n_5$ is not a strictly conserved quantity. So studies on chiral imbalance effect is considered as  on a time scale much larger than the typical time scale of the chirality changing processes \cite{Ruggieri16}. And the possible impact of chiral imbalance on the EOS is neglected on the literature.  But chiral density may induce by electro-magnetic fields. As showed in the Eq. (1) in  Ref. \cite{Ruggieri16} with parallel magnetic fields in the background,  the chiral density $n_5$ is proportional to the magnetic field strength. We known that pulsars  are found with strong magnetic field. Thus chiral imbalance  has a greater probability of occurrence in pulsars and CME will be more obvious.
In this paper, we will firstly study the effect of chiral imbalance on the equation of state (EOS) under proper-time regularization. Furthermore, the influence on the mass-radius relations and tidal deformability of quark stars will be investigated and hope to find some clue  for CP  violation in compact object.

In this paper, the chiral phase transition is studied under  the recent developed NJL model with proper-time regularization \cite{wangqy,zhaot,wangqw}.  The  standard two-flavor Nambu$-$Jona-Lasinio (NJL) Lagrangian  contains only scalar and pseudoscalar-isovector channels. But its Fierz transformation, as a   mathematically equivalence,  contains more interactive channels \cite{Klevansky}, especially the vector channel.
Except the chiral chemical potential, model calculations show that the vector   channel  will also  affect chiral phase structure and the location of CEP \cite{Gatto,Wangf}.  The critical chemical potential will increase as the vector  coupling increases. When the  coupling is larger enough, the CEP will disappear.  The contribution from the vector channel  is quite important at nonzero densities.  The strength of the vector coupling   is usually taken as free parameter in quark model. In the relativistic mean-field model of nuclear matter, the vector couplings are fitted to low-energy data.    But whether it would be suppressed at high temperature and high density is unknown. Then   density-dependent couplings are proposed in the exploring of hot and dense nuclear matter.  Therefore, the detection of CEP in heavy-ion collisions will also provide information about vector  channel interactions.

The linear combination of standard NJL Lagrangian and  its Fierz transformation will include the vector  channels interaction in a consistent way. The concerned channels in the Fierz transformation are the vector channel $-(\bar\psi\gamma^\mu\psi)^2$ and the axial-vector channel $-(\bar\psi i\gamma_5\gamma^{\mu}\psi)^2$.  In the mean-field-approximation,
\begin{eqnarray}
  -(\bar\psi\gamma^\mu\psi)^2 & \approx&-2n \psi^\dagger \psi  +n^2,  \label{eq.1} \\
 -(\bar\psi i\gamma_5\gamma^\mu\psi)^2 &  \approx& 2 n_5\psi^\dagger \gamma_5\psi-n_5^2. \label{eq.2}
\end{eqnarray}
Here, the $n$ and $n_5$ are the  number density and the chiral number density of quarks respectively.

 This paper is organized as follows: In Sec. II, we introduce the  newly developed self-consistent mean-field theory of the NJL model.    In Sec. III, We give our numerical results and analysis on the phase transition. Sec. IV is a short summary of our work.

\section{ Nambu$-$Jona-Lasinio model}

 In the recently developed self-consistent two-flavor NJL model \cite{wangqy,zhaot,wangqw}, the Lagrangian can be expressed as a linear combination of a standard NJL Lagrangian ($\mathcal{L}_{NJL}$) and its Fierz transformation ($\mathcal{L}_{Fierz}$) \cite{Buballa,Klevansky,Kunihiro}, that is

 \begin{eqnarray}
\mathcal{L} _C=\left(1-\alpha \right)\mathcal{L}_{NJL} + \alpha \mathcal{L} _{Fierz},
\label{eq.la}\end{eqnarray}
where $\alpha$   weights the contribution from Fierz transformation.
At finite density,   $\mu   \psi^\dagger  \psi$ can be added to the right of Eq. \eqref{eq.la}.  Similarly, we can consider quark chiral imbalance  by adding a term  $\mu_5  \psi^\dagger  \gamma_5\psi$ to the right side with $\mu_5$ being the chiral chemical potential coupling to the chiral operator.
Here, we use the standard NJL  Lagrangian with four-quark interaction for $\mathcal{L}_{NJL}$. The $\mathcal{L}_{NJL}$ and its Fierz transformation $\mathcal{L}_{Fierz}$ can be written, respectively,  as

  \begin{eqnarray}
\mathcal{L}_{NJL}=\bar{\psi }(i\slashed{\partial } - m)\psi + G[\left(\bar{\psi }\psi\right)^2+\left(\bar{\psi } i \gamma _5 \vec{\tau }\psi \right)^2],
\end{eqnarray}
 and
\begin{eqnarray}
\begin{aligned}
\mathcal{L} _{Fierz}=& \bar{\psi }(i\slashed{\partial } - m)\psi +  \frac{G} {8 N_c}[2\left(\bar{\psi } \psi \right)^2+2\left(\bar{\psi } i \gamma _5 \vec{\tau} \psi \right)^2\\
& - 2\left(\bar{\psi } \vec{\tau} \psi \right)^2 - 2\left(\bar{\psi } i \gamma _5 \psi \right)^2 - 4\left(\bar{\psi } \gamma^{\mu } \psi \right)^2\\
& - 4\left(\bar{\psi } i\gamma^{\mu } \gamma_5\psi \right)^2 + \left(\bar{\psi } \sigma^{\mu \nu} \psi \right)^2 - \left(\bar{\psi } \sigma^{\mu \nu} \vec{\tau} \psi \right)^2].
\end{aligned}
\end{eqnarray}
Under the mean-field approximation, the effective quark mass is
  \begin{eqnarray}\label{eq.muc2}
    M&=&m-2G^\prime\sigma.
     \end{eqnarray}
Here, $\sigma=\left\langle{\bar\psi  } \psi\right\rangle$ is the two-quark condensation and  $G^\prime$ is the four-quark effective coupling for the mixed Lagrangian Eq. (\ref{eq.la}) which has the relation with $G$
\begin{equation}\label{eq.gprime}
  G^\prime=(1-\alpha+\frac{\alpha}{4N_c})G.
\end{equation}
It is the new coupling $G^\prime$ that  needs to be recalibrated to fit the low-energy experimental data.
The modified chemical potential and chiral chemical potential are defined, respectively, as
  \begin{eqnarray}\label{eq.muc2}
    \mu_r&=&\mu- \frac{\alpha G} {  N_c  }n, \label{eq.mu}\\
   \mu_{5r}&=&\mu_5+ \frac{\alpha G} {  N_c  }n_5. \label{eq.mu5}
  \end{eqnarray}
Here, $n=\left\langle{\psi ^\dagger } \psi\right\rangle$ is the quark number density and $n_5=\left\langle{\psi ^\dagger } \gamma_5\psi\right\rangle$ is the chiral number density.  Note here that,  because the introduction of the four-quark pseudo-vector interaction are different  by a negative sign here and in Eq. (31) of Ref.  \cite{Ruggieri}, the modified $\mu_5$ differs by a negative sign in Eq. \eqref{eq.mu5} and Eq. (34) of Ref.  \cite{Ruggieri}.

The chiral condensate and (chiral-) quark number densities are given by minimizing the thermodynamic potential density. At finite density and temperature, the chiral condensate is
  \begin{eqnarray}\label{eq.gapt0}
\sigma&=&-\frac{N_cN_f M T}{2\pi ^2} \sum_{s=\pm 1}\sum^\infty_{n=-\infty}\int  \frac{ \mathbf{ p}^2}{E_s^2+\tilde \omega_n^2} dp ,
\end{eqnarray}
where $\tilde \omega_n$ is the fermion Matsubara frequency which is defined as  $\tilde \omega_n=   \omega_n+i \mu$, $ \omega_n= (2n+1)\pi T$ with  $n \in  \mathbb{Z}$.  And the energy for different helicity $s$ is defined as $E_s$=$\sqrt{M^2+(\mu_{5r}-s|\mathbf{p}|)^2}$ with $s=\pm 1$.
In the proper-time regularization scheme, $1/A(p^2)$ is replaced by $ \int_{\tau_{UV}}^ \infty d\tau e^{-\tau A(p^2)}$,  with $\tau_{UV}=1/\Lambda^2_{UV}$  and $\Lambda_{UV}$ is the UV cutoff to regularize the ultraviolet divergence.  The chiral condensate at finite temperature and density can be written as in Ref. \cite{Cui} as

\begin{eqnarray}\label{eq.gapt}
\sigma&=&-\frac{N_cN_f M }{2\pi ^2} \sum_{s=\pm 1}\int_0^\infty  \int_{\tau_{ UV}}^\infty \frac{p^2 e^{-\tau E_s^2}}{ \sqrt{\pi\tau}}\times\\
&& [1-f_s^-(p,\mu_r,\mu_{5r},T)-f_s^+(p,\mu_r,\mu_{5r},T)]d\tau dp, \nonumber\\
&=&-\frac{N_cN_f M }{2\pi ^2} \sum_{s=\pm 1}\int_0^\infty \frac{p^2}{E_s}\texttt{Erfc}(\sqrt{\tau_{UV}}E_s)\times \\
&& [1-f_s^-(p,\mu_r,\mu_{5r},T)-f_s^+(p,\mu_r,\mu_{5r},T)] dp . \nonumber
\end{eqnarray}
 Here Erfc$(x)$ is the complementary error function and
  the $f_s^\pm$ defines the Fermi-Dirac distribution function under the modified (chiral-) chemical potentials and nonzero temperature $T$, with
 \begin{eqnarray}\label{eq.gapt}
  f_s^\pm&=& \frac{1}{1+e^{(\omega_s\pm\mu_r)/T}} .
\end{eqnarray}
Similarly, the quark number density $n$ and the chiral number density $n_5$ are

\begin{eqnarray}\label{eq.gapt}
n&=&\frac{N_cN_f }{2\pi ^2} \sum_{s=\pm 1}\int_0^\infty p^2  [f_s^-(p,\mu_r,\mu_{5r},T) \nonumber\\
&&-f_s^+(p,\mu_r,\mu_{5r},T)] dp, \\
n_5&=&\frac{N_cN_f }{2\pi ^2} \sum_{s=\pm 1}\int_0^\infty  p^2\frac{\mu_{5r}-sp}{E_s}\texttt{Erfc}(\sqrt{\tau_{UV}}E_s) \nonumber \\
&& [1-f_s^-(p,\mu_r,\mu_{5r},T)-f_s^+(p,\mu_r,\mu_{5r},T)]dp .
\end{eqnarray}

With $f_\pi=93$ MeV,  $m_\pi= 135$ MeV, and $m = 3.5$ MeV, three parameters ($m$,   $G^{\prime}$,  and $\tau_{UV}$)  are fixed  to fit the Gell-Mann$-$Oakes$-$Renner relation:$- 2 m \left\langle {  \bar \psi \psi} \right\rangle$ =  $(f_\pi m_\pi)^2$.
The quark condensate is $\left\langle {  \bar \psi \psi} \right\rangle)^{1/3}=-282.4$ MeV. Then we have  $G^\prime=4.1433 \times 10^{-6} $MeV$^{-2}$  and $\Lambda_{UV}$ =  955   MeV. The coupling $G$ is adjusted with $\alpha$.

In order to study the response of chiral condensate to chemical potentials and temperature, the susceptibilities are defined by
\begin{eqnarray}
\chi_\mu=-\frac{\partial \sigma}{\partial\mu}, \quad \chi_{\mu_5}=\frac{\partial \sigma}{\partial\mu_5}, \quad \texttt{and}\quad \chi_T=-\frac{\partial \sigma}{\partial T}.
\end{eqnarray}

\begin{figure}[h]
    \includegraphics[width=0.9\columnwidth]{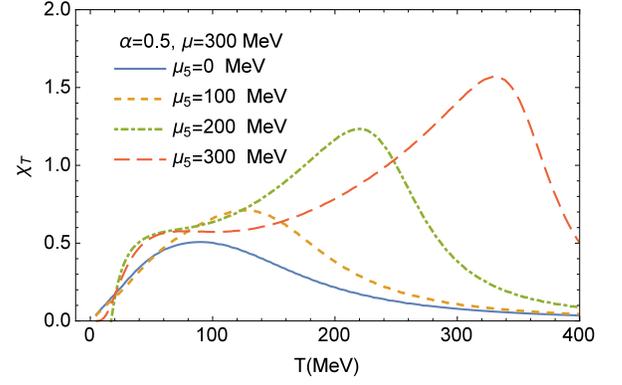}
    \caption{The temperature susceptibility  for different  chiral chemical potential $\mu_5$. } \label{fig.chitmu300}
\end{figure}

\section{results and analysis}

 For different $\alpha$, the pseudo-transition point is around $\mu=300$ MeV at $T=0$ and increases as $\alpha$ increases which is similar to the results from three-momentum cutoff  scheme. For nonzero temperature, the $\chi_T$ is showed in Figure \ref{fig.chitmu300}.  It is crossover even when $\mu=300$ MeV.
 It is different from the three-momentum cutoff  scheme that the proper-time regularization scheme gives a crossover for the chiral phase transition even at vary large $\mu$  and thus   CEP does not exist in the chirally balanced system ($\mu_5=0$) in this regularization scheme. We will show the results with $\mu_5 \neq 0$ bellow.

 \begin{figure}[h]
    \includegraphics[width=0.9\columnwidth]{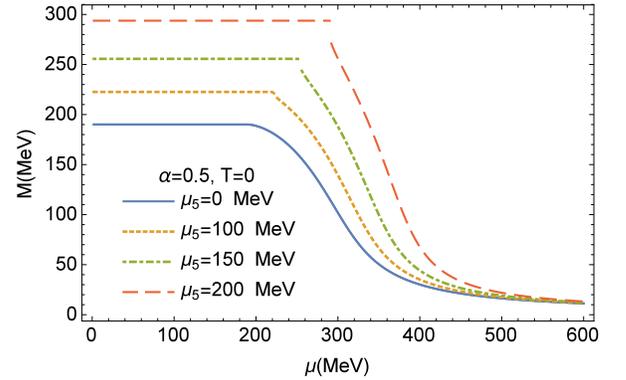}
    \caption{The quark mass as a function of chemical potential $\mu$ for different  chiral chemical potential $\mu_5$.   }\label{fig.mq}
\end{figure}
\subsection{ The exists of CEP}

As chiral imbalance appeared, a jump down in the $M-\mu$ plot   appears, as showed in Figure \ref{fig.mq}. The jump occurs at about $\mu=M_{vac}$ only for relatively large $\mu_5$, where $M_{vac}$ is the quark mass at $T=\mu=0$. Beside, it shows that quark mass  increases   for different chiral chemical potential. The phenomenon that chiral condensate  increases with some external field is called  `catalysis'.  Similar to the magnetic catalysis \cite{Elmfors, Persson, Ebert, Inagaki}, it may be called as chiral catalysis here since it is caused by chiral imbalance. The inverse magnetic catalysis is   found \cite{Bali2012,Wang2016}.  Here in the proper-time scheme, only  chiral catalysis  exists with constant couplings.   In the three-momentum cutoff  scheme,   inverse chiral catalysis exists as the chiral symmetry is partly restored \cite{Luya}.

\begin{figure}[h]
\subfigure[]{    \includegraphics[width=0.45\columnwidth]{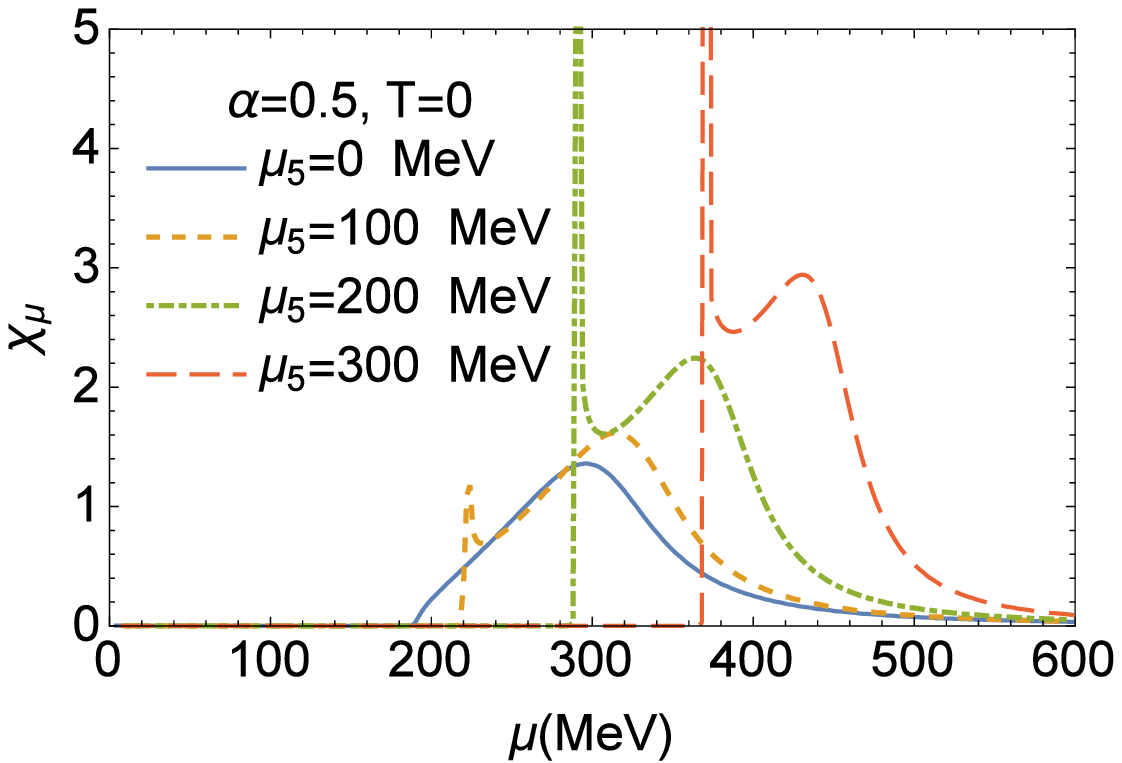}  \label{fig.chimu1}}
 \subfigure[]{ \includegraphics[width=0.45\columnwidth]{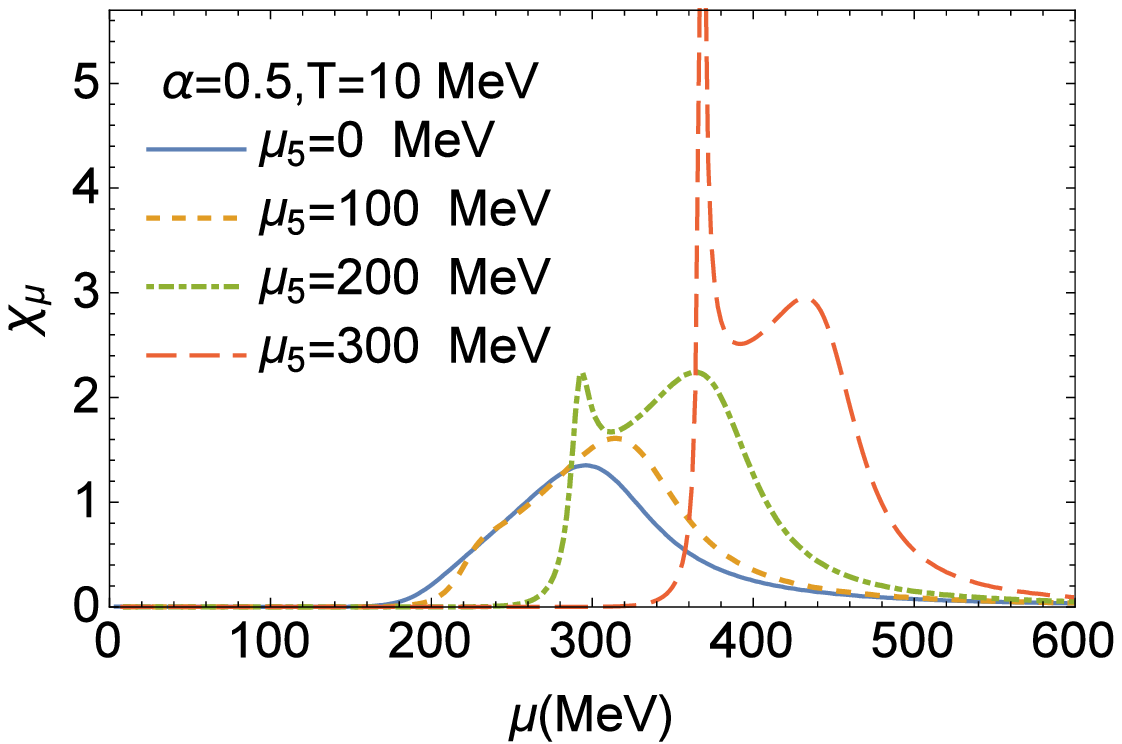}\label{fig.chimu2}}\\
  \subfigure[]{  \includegraphics[width=0.45\columnwidth]{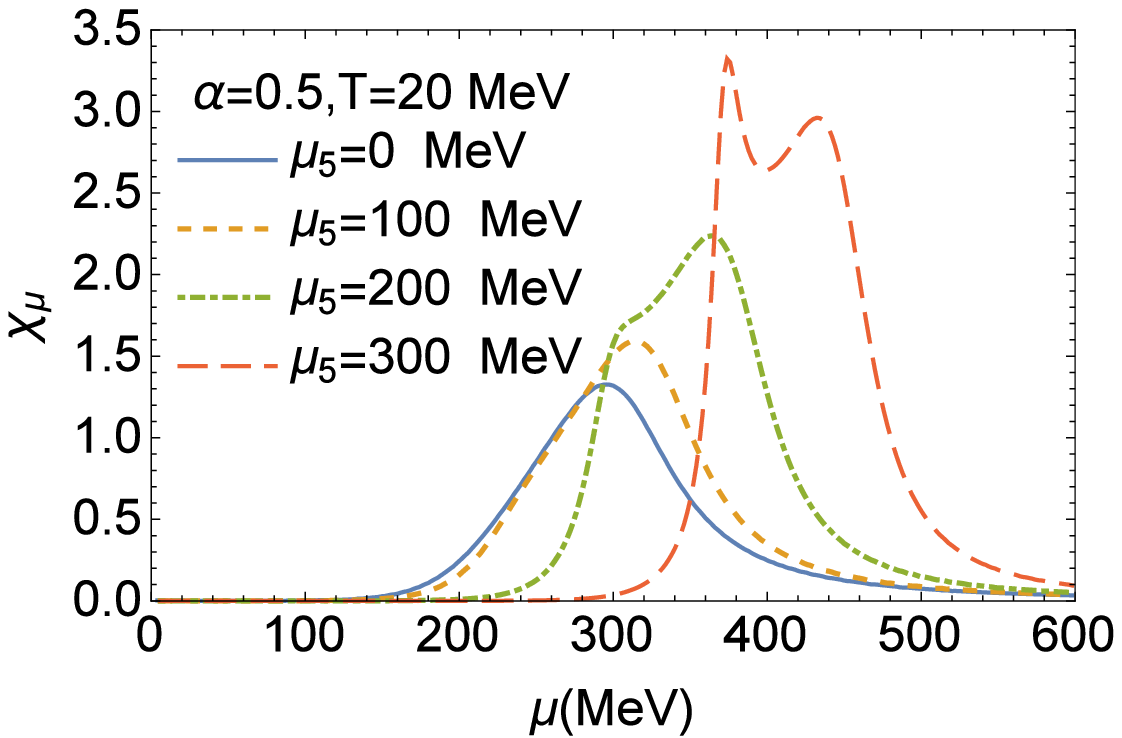}\label{fig.chimu3}}  \subfigure[]{  \includegraphics[width=0.45\columnwidth]{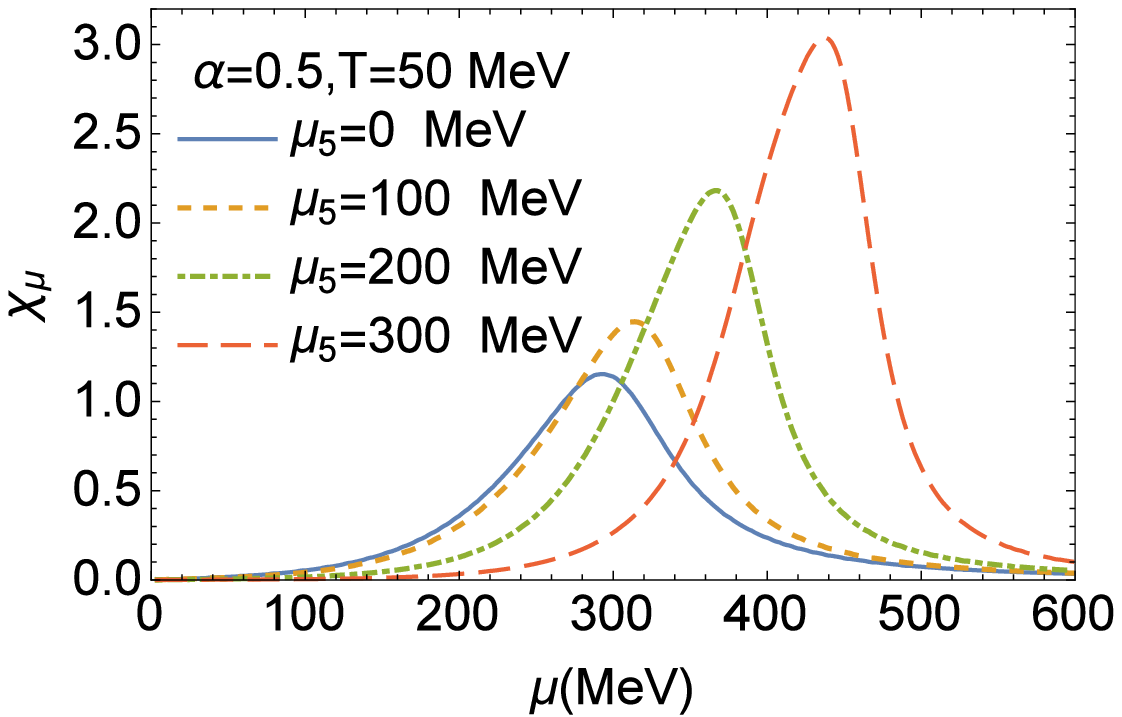}\label{fig.chimu4}}
    \caption{The susceptibilities as a function of $\mu$ for different chiral chemical potential $\mu_5$ and temperature $T$.  } \label{fig.chimu}
\end{figure}

In Figure \ref{fig.chimu}, the chemical susceptibilities $\chi_\mu$ for different $\mu_5$ at T=0,10,20, and 50 MeV are presented.
When $\mu_5=0$, the chiral transition is always a crossover at any $T$ and $\mu$. As $\mu_5$ is nonzero, peaks and bumps appear in the plots.  When $T=0$ and  $\mu_5>0$, the increasing of $\mu$ will cause significant peaks in the  susceptibility lines.  This is a phenomenon different to results when $\mu_5=0$  which  strongly suggests that there is a first-order phase transition.    It is obvious that   the   critical chemical potentials increase with $\mu_5$.
As temperature increases, the peaks begin to disappear even for nonzero $\mu_5$ and remains only pumps. This change may indicate the existence of CEP in chiral imbalance system.

\subsection{The exists of CEP$_5$}

\begin{figure}[h]
    \includegraphics[width=0.9\columnwidth]{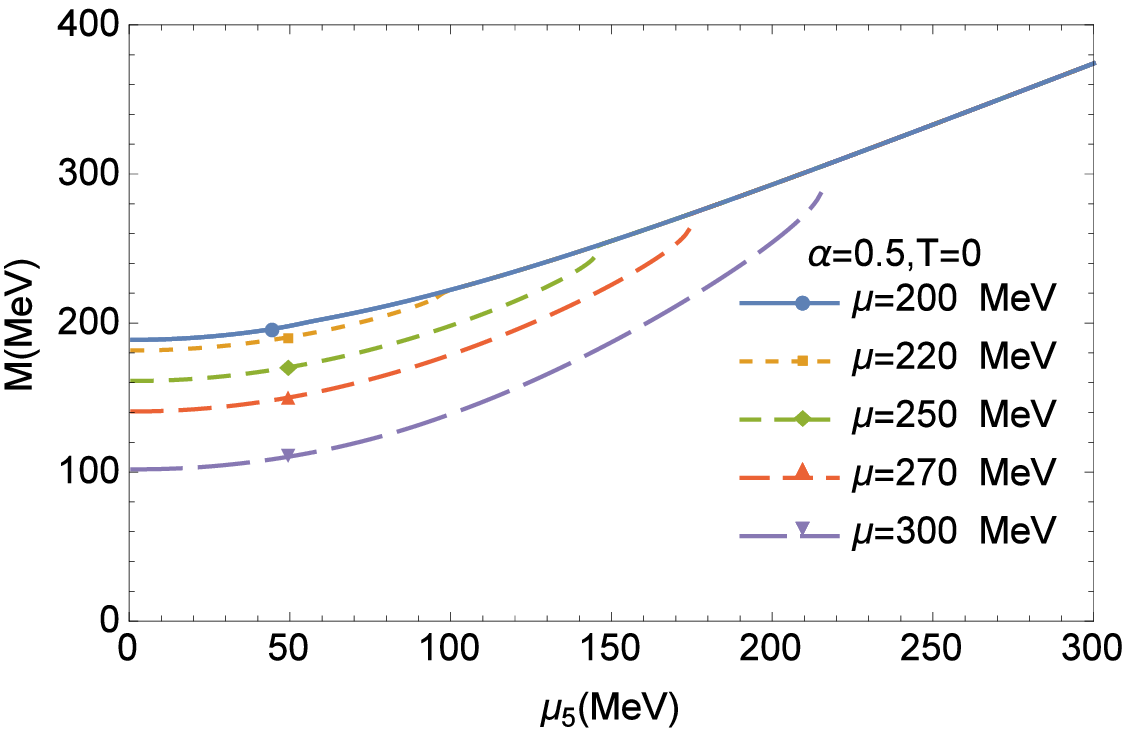}\\
 \includegraphics[width=0.9\columnwidth]{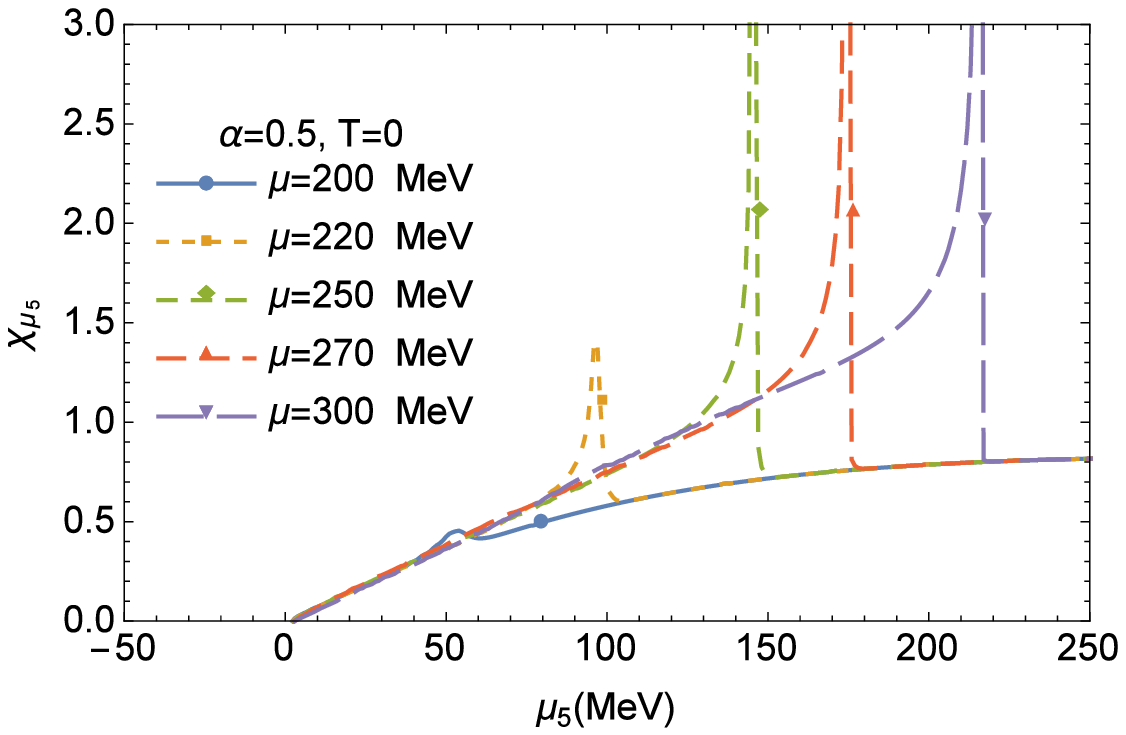}
    \caption{Chiral catalysis effect and chiral susceptibility at zero temperature. In the up-plane, as $\mu_5$ is larger than some critical values, lines of $\mu>200$ MeV  coincide with line of $\mu=200$ MeV .    } \label{fig.chimu5}
\end{figure}

In order to show the influence of chiral imbalance more clearly, we show the quark mass as a function of chiral chemical potential $\mu_5$ in Figure \ref{fig.chimu5}. And the corresponding susceptibilities  for fixed baryon chemical potentials are also presented.

As chemical potential $\mu$ less than 200 MeV, the quark mass almost unchanged with $\mu$ but smoothly increases with chiral chemical potential $\mu_5$. So, the phase diagram in the $T-\mu_5$ plane is crossover in this region of $\mu$.
   As $\mu \geq 220$ MeV, obvious peaks are showed up in the plot of $\chi_{\mu_5}$ which implies the exists of the  first order phase transition and  CEP$_5$.
      The plot also shows  that even if the CEP$_5$ exists, there is a threshold for chemical potential $\mu$. However, since the quark condensate or mass does not change obviously here at the transition point, it is hard to find exactly where the first order phase transition begins.   At T=0, the beginning of first order transition  in the $\mu_5-\mu$ plane is located in regions of  $\mu\in (200,220)$ MeV and close to $\mu_5=100$ MeV.

\begin{figure}[h]
    \includegraphics[width=0.9\columnwidth]{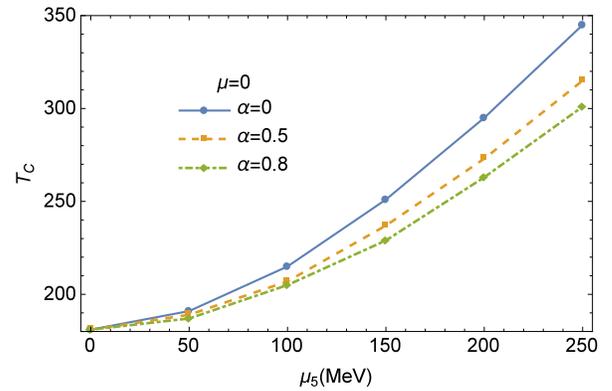}
    \caption{The  pseudo-  critical temperature $T_c$ as a function of chiral chemical potential $\mu_5$  } \label{fig.p1}
\end{figure}

\begin{figure}[h]
    \includegraphics[width=0.9\columnwidth]{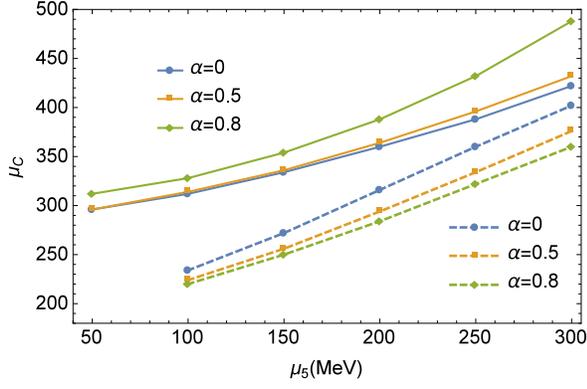}
    \caption{The (pseudo-) critical chemical potential $\mu_c$ as a function of chiral chemical potential $\mu_5$. The solid  lines show the pseudo critical chemical potential   while the dashed lines show the critical chemical potential. } \label{fig.p2}
\end{figure}

The critical temperature in the CEP plane will increase with temperature. Since it is also hard to locate the CEP, to find a relation of CEP and CEP$_5$ seems not possible in the proper-time regularization scheme.
Here, we show the pseudo-critical temperature $T_c$ and  critical chemical potential $\mu_c$ (T=0) as a function of chiral chemical potential $\mu_5$   in Figure \ref{fig.p1} and Figure \ref{fig.p2}, respectively.
 As $\mu=\mu_5=0$, the pseudo-critical temperature is about 181 MeV. The $T_c$ decreases with $\alpha$ which is similar to the result with three-momentum cutoff regularization. But the $T_c$ and $\mu_c$   increase  with $\mu_5$. This result is different from the result with three-momentum cutoff regularization. In Figure \ref{fig.p2}, both pseudo-critical and  critical chemical potentials as a function of $\mu_5$ at T=0 are presented. The solid lines represent the pseudo-    critical chemical potentials which are larger than the critical chemical potentials (dotted lines). No critical chemical potentials exist at small $\mu_5$.The pseudo-    critical chemical potentials increases with $\alpha$, while  the     critical chemical potentials increase with $\alpha$. This trend can also be seen from Figure \ref{fig.chimu1}. With the increase of $\mu_5$, the peak and bump become close to each other. For very large $\mu_5$, pump disappears and only  critical chemical potentials exists.

\subsection{The EOS of quark matter in chirally imbalanced system}
\begin{figure}[h]
    \includegraphics[width=0.9\columnwidth]{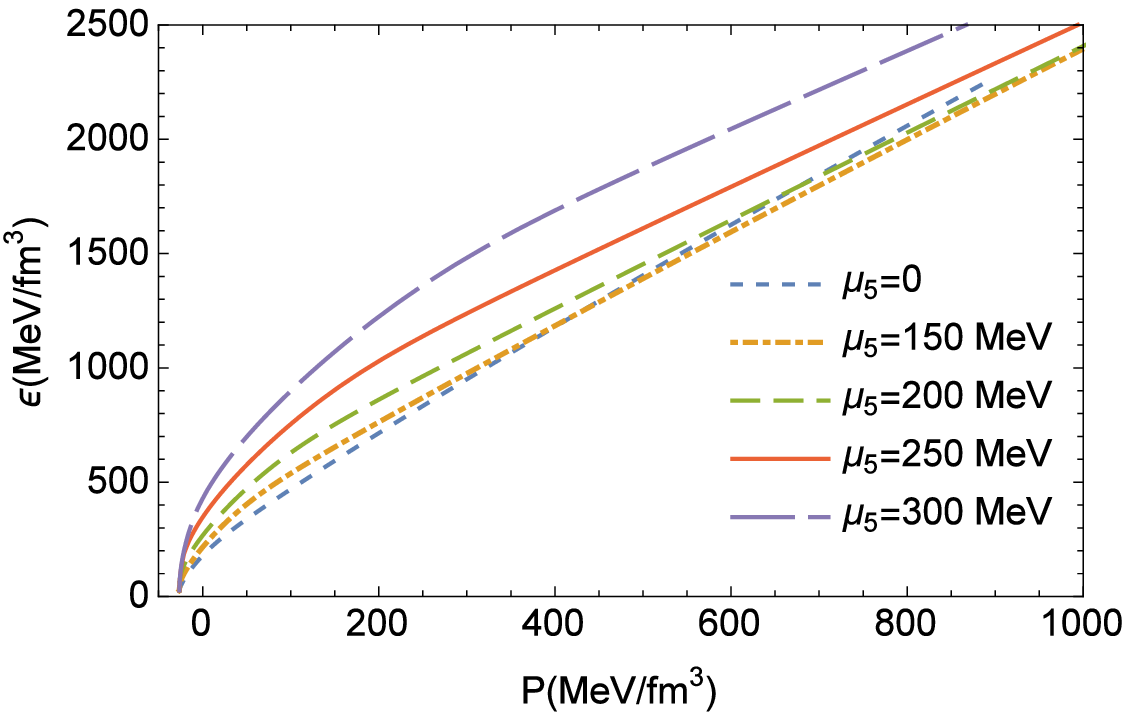} \\
     \includegraphics[width=0.86\columnwidth]{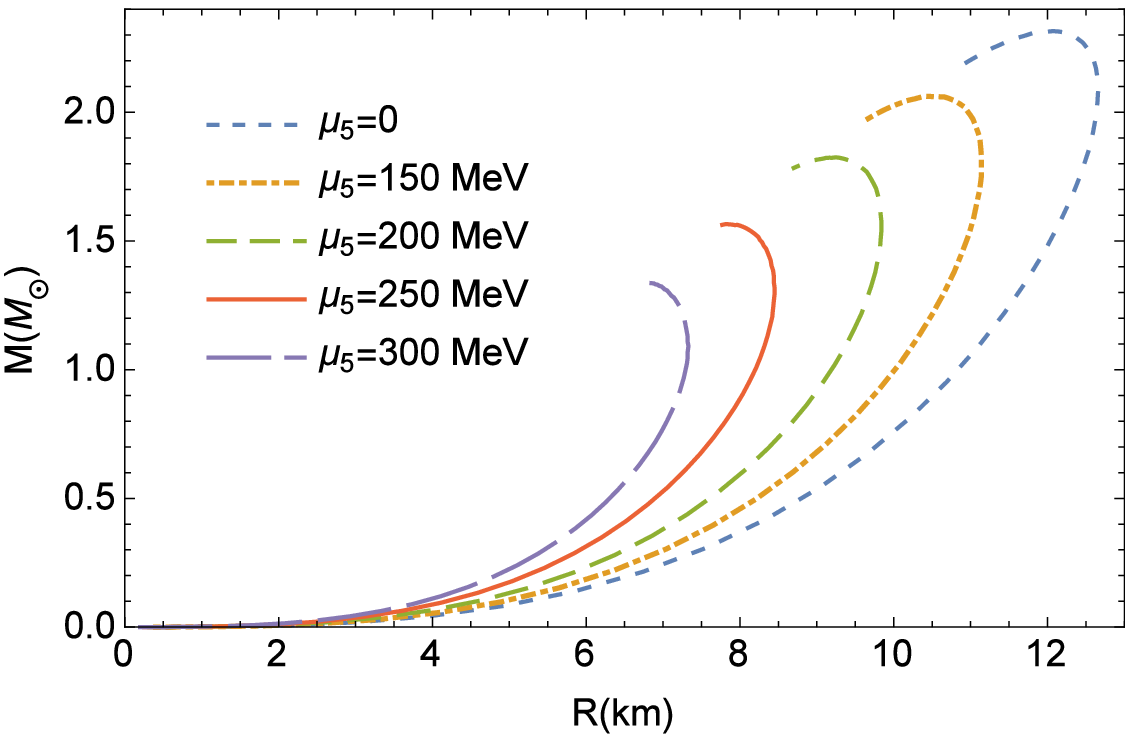}
    \caption{The influences of chiral imbalance on the EOS of quark matter and the mass-radius relation of   quark stars. The tidal deformabilities are 827.237, 421.616, 194.621, and  56.618,  respectively, which decreases as $\mu_5$ increases.
    } \label{fig.eos}
\end{figure}

In the end we study the influence of chiral imbalance on the EOS of quark matter and mass-radius relation of quark stars. If first-order phase transition were found in the astronomical observation,
 it will confirm the exist of CEP \cite{Blaschke, Alvarez} and result in the emergence of a third family of compact stars, in addition to white dwarfs and neutron stars \cite{Gerlach,Kaltenborn,Benic}.
 For quark matter, the model-independent  equations of state  of strong interaction matter  are  \cite{zong1,zong2}
\begin{eqnarray}
P(\mu)&=&P_0 +  \int_ 0^{\mu }d\mu n  (\mu ), \label{eq.p}\\
\epsilon(\mu)&=&-P(\mu)+ \mu n(\mu).\label{eq.e}
\end{eqnarray}
Here, $P_0$ represents the vacuum pressure  at $\mu=0$ and is taken as $P_0=-(120$ MeV)$^4$. The mass of recent observed pulsars \cite{Demorest,Antoniadis,Fonseca,Cromartie} is about $2M_{\odot}$.  We set $\alpha=0.8$ to ensure the EOS is stiff enough.
In studying the mass-radius relation,  we use  the  static TOV equations
(in units $G=c=1$ )
\begin{eqnarray}
 \frac {dP\left(r \right)} {dr} &=&-\frac {\left(\epsilon +P \right)\left(M + 4\pi r^3 P \right)} {r \left(r-2M \right)},\\
\frac {dM\left(r \right)} {dr} &=&4\pi r^2 \epsilon.
 \end{eqnarray}
Here, $P$ and $\epsilon$ are the pressure and energy density as in Eqs. \eqref{eq.p} and   \eqref{eq.e}. $M(r)$ is the quark star mass as a function of radius  $r$.  The equations are solved iteratively from a central pressure to zero pressure that defines the edge of the star \cite{Glendenning}.
Figure \ref{fig.eos} shows that the EOS becomes soft as $\mu_5$ increases. The maximum masses and radii of quark star  reduce as $\mu_5$ increases. When $\mu_5$ is greater than $150$ MeV, the maximum mass   becomes less than 2  $M_\odot$. When $\mu_5$ is greater than $200$ MeV, the maximum radius   becomes less than 10   km. If the observed pulsars of masses larger than 2 $M_\odot$ are identified as quark stars and radius larger than 10  km,  then the chiral chemical potential cannot be very large.

 During the merger of two compact stars, the  tidal formability $\Lambda$  measures the star's quadrupole deformation in response to the companion's perturbing tidal field.
The  tidal deformability  can be expressed through   compactness $C= M/R$ and Love number $k_2$. Here, $M$ and $R$  are the star mass and radius, respectively. The relation is

\begin{equation}\label{eq.lambda}
k_2=\frac{3}{2}\Lambda\left( \frac{M}{R}\right)^5.
\end{equation}
 The method to calculate $k_2$  can be found in Refs. \cite{Damour,wangqw,licm2020}.
 The calculated results show that with the increase of chiral chemical potentials the tidal deformability decreases.

\subsection{ Comparison of the influence of vector channel and axial-vector channel }

\begin{figure}[h]
    \includegraphics[width=0.86\columnwidth]{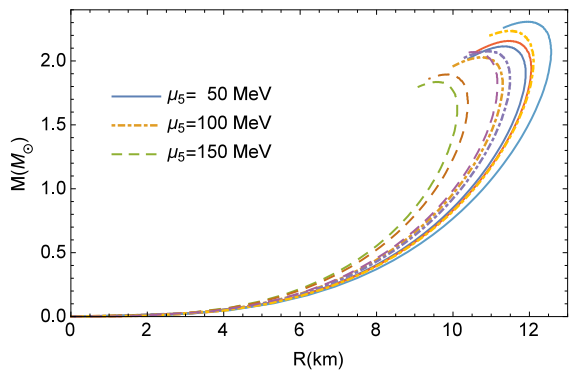}

    \caption{ Mass-radius relations for different $\alpha$ and $\mu_5$.  When   $\mu_5$ is fixed, the maximum mass increases with $\alpha$. Lines with the same $\mu_5$   are corresponding to $\alpha=$0.5, 0.6, and 0.8, respectively.  } \label{fig.mr9}
\end{figure}

 The increase of weight factor $\alpha$ and chiral chemical potential $\mu_5$ has opposite effect on the stiffness of EOS. With the increase of $\alpha$, the EOS will become hard, whereas with the increase of $\mu_5$, EOS will become soft. Therefore, the increase of $\mu_5$ will reduce the maximum mass in the mass-radius plot. A comparison of the mass-radius relation with different $\alpha$ and $\mu_5$ is showed in Fig. \ref{fig.mr9}  and the tidal deformabilities for a 1.4 $M_\odot$ quark star are listed in Table \ref{tab.mr9}.

  \begin{table}
\centering
\caption{The tidal deformability of 1.4 $M_\odot$ quark star for different $\alpha$ and $\mu_5$(MeV) .  }	\label{tab.mr9}
\begin{tabular}{l|ccccccc}
\hline \hline
\diagbox{$\alpha$} {$\mu_5$} & 50&  100&150    \\\hline
0.5&606.198& 450.802& 226.158    \\
0.6 &648.325&497.448&271.404  \\
0.8& 790.556 & 654.638&421.616 \\
 \hline\hline
\end{tabular}
\end{table}

\begin{figure}[h]
    \includegraphics[width=0.86\columnwidth]{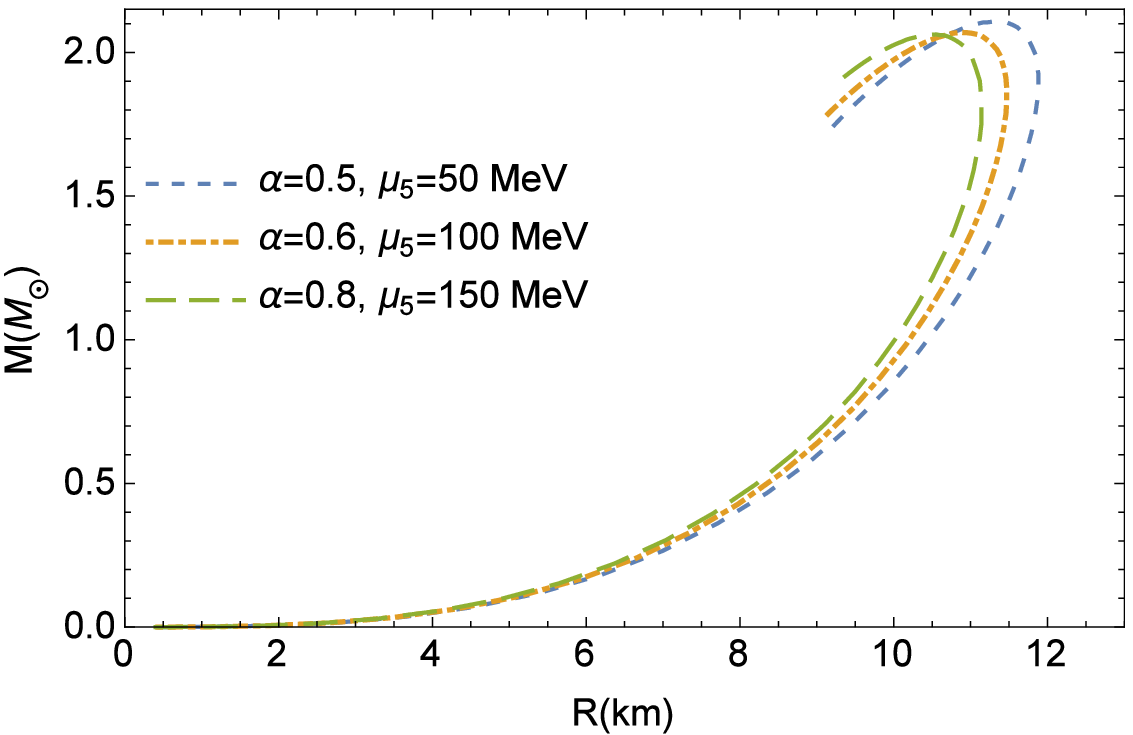} \\
     \includegraphics[width=0.9\columnwidth]{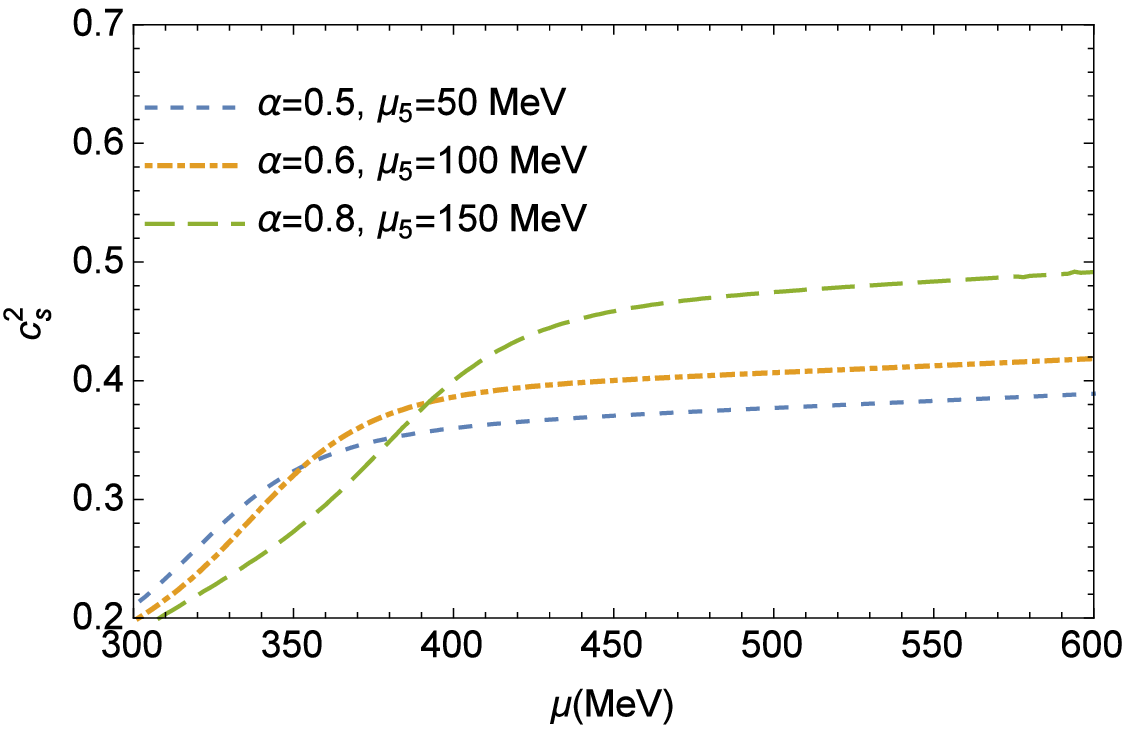}
    \caption{ Mass-radius relations and squared speed of sound. The mass-radius lines are setting to have almost equal maximum but the  mass-radius relations and squared speed of sound $c_s^2$ are different which can be identified indirectly as a signal of CP violation inside the compact stars. } \label{fig.mr2}
\end{figure}

  In addition,the increase of $\alpha$ will lead to the phase transition from first-order to crossover, whereas $\mu_5$ will lead to the transition from crossover to first-order. It seems that the effects of the two parameters may cancel each other out. When we think about them at the same time, we can not tell which factor (vector channel and chiral   imbalance) plays the main role.
However, the existence of $\mu_5$ may affect the shape of mass-radius relationship. That is to say, without considering other external parameters, but constraining EOS by the maximum mass or  tidal deformability from astronomical observation, the shape of mass-radius curves by including  vector channel and axial-vector channel may be different from  the one only considering vector channels.
 As can be seen from Fig. \ref{fig.mr2}, our calculations confirm this. When the maximum masses are limited to about 2 $M_\odot$, the radii corresponding to the maximum mass and the mass-radius curves are still different. The introducing of chiral chemical potential can make the tidal deformability have a right  value. If we restrict the tidal deformability for a 1.4 $M_\odot$ star to the range 70$<\Lambda(1.4 M_\odot)<$580 \cite{abbott2, annala20}, the calculated tidal deformabilities with large $\alpha$ and small $\mu_5$ is out of the range.
  That is to say, for quark stars when other parameters (such as magnetic field strength, rotation speed, temperature) are well understood, it is possible to observe the effectiveness of chiral chemical potential by analyzing the mass-radius relation and tidal deformability,   so as to indirectly prove the violation of strong CP in compact stars.

\section{summary }

The chiral imbalance indicates the nonequal density  of left- and right- quarks that may occur in the QGP phase. In this paper,  we study the influence of   chiral chemical potential $\mu_5$ on the chiral   phase transition and the EOS of quark matter.   We use two-flavor NJL model with  proper-time regularisation.  And we know that there is no first-order phase transition at $\mu_5=0$ in this regularisation scheme. When $\mu_5$ increases, the first-order phase transition appears. But the phase transition is not very strong. So it is difficult to find the position of CEP.  On the other hand with fixed $\mu$, the chiral phase transition   responding to chiral chemical potential $\mu_5$ is found to be a first-order phase transition, ie., the CEP$_5$ exists.

 We have calculated the   pseudocritical temperature  $T_c$ and (pseudo-)critical chemical potentials $\mu_c$ as a function of $ \mu_5$. It is found that these quantities increase rather than decrease with $\mu_5$. This is different to the result from three-momentum cutoff schemes. Besides, there are two maxima for the chiral susceptibility $\chi_{\mu} $ at zero temperature.
 With the increase of $\mu_5$, the two maxima gradually approach   to each other. In addition, we have also calculated the chiral susceptibility $\chi_{\mu} $ at different temperatures. It is found that with the increase of temperature, the peaks gradually disappear and only the bumps remain. This further confirmed the existence of CEP when $\mu_5$ is not zero.  At last, we have calculated the EOS of quark matter and the mass-radius relation  at different $ \mu_5$.   The calculations show that the EOS becomes soft with the increase of $\mu_5$. When parameters are set so that  the maximum mass of quark stars is larger than $2  M_\odot$, it is found that with the increase of $\mu_5$, the maximum mass decreases gradually to masses far less than $2 M_\odot$ and the maximum radius  is less than $10$ km .

We have also showed that when the  mass-radius relations are constrained by maximum mass, the different radius that corresponding to the maximum mass and the variety of the mass-radius cures can be used to as a confirmation of  the existence of chiral imbalance   and indirectly  prove the violation of strong CP in dense matter. However, this depends on the accurate measurement of the radius of  compact star.

  In  conclusion, the chiral imbalance has significant influence on  the chiral phase structure of quark matter. Since the chiral imbalance may possibly exist in the QGP phase, it is important to consider it when confirm and locate the position of CEP in the experiments.

\acknowledgments
This work is supported in part by the National Natural Science Foundation of China (under Grants No. 11475085,  No. 11535005, No. 11690030, No.11873030, and No. 11905104) and the National Major state Basic Research and Development of China (Grant No. 2016YFE0129300).

\end{document}